\documentclass[a4paper,10pt,openany]{article}
\usepackage{pst-3d}
\usepackage{pst-3dplot}
\usepackage{etex}
\usepackage[utf8]{inputenc}
\usepackage[english]{babel}
\usepackage{amsmath,amssymb,amsthm,mathrsfs,amsfonts,dsfont}
\usepackage{graphicx}
\usepackage[footnotesize, skip=0pt]{caption}
\usepackage{epigraph}
\usepackage{indentfirst}
\usepackage{booktabs}
\usepackage{fancyhdr}
\usepackage{vmargin}
\usepackage{feynmf}
\usepackage{yfonts}
\usepackage{lscape}
\usepackage{subfig}
\usepackage{textcomp}
\usepackage{pstricks}
\usepackage[metapost]{mfpic}
\usepackage{fancybox}
\usepackage{slashed}
\usepackage[verbose]{wrapfig}
\usepackage{pifont}
\usepackage{keystroke}
\usepackage{appendix}
\usepackage{enumitem}
\usepackage{array}
\usepackage{colortbl}
\usepackage{alltt}
\usepackage{boxedminipage}
\usepackage{tikz}
\usepackage{calc}
\usepackage{subfloat}
\usepackage{multirow}
\usepackage{cmll}
\usepackage{multicol}
\definecolor{color1}{RGB}{204,0,51}
\definecolor{color2}{RGB}{159,182,205}
\usepackage{bookmark}

\usepackage{verbatim}
\usepackage[vcentermath]{youngtab}
\usepackage{young}
\usepackage{pst-grad} 
\usepackage{pst-plot}
\usepackage{transparent}
\usepackage{color,soul}
\usepackage{cancel}
\usepackage{bm}
\usepackage{pifont}
\usepackage{placeins}
\usepackage{mathtools}
\usepackage{rotating}
\usepackage[refpage]{nomencl}
\usepackage{hhline}
\usepackage{marginnote}
\usepackage{upgreek}
\usepackage{stackengine}
\usepackage{setspace}
\usepackage{ytableau}
\usepackage{listings}
\usepackage{calligra}
\usepackage{hyperref}
\usepackage{afterpage}
\usepackage{cleveref}
\usepackage{makecell}
\usepackage{accents}
\usepackage{breqn}
\usepackage{environ}

\numberwithin{equation}{section}

\newcommand{\EAB}[1]{{\color{black}#1}}

\allowdisplaybreaks

\sethlcolor{yellow}
\definecolor{boh}{RGB}{79,47,79}

\hypersetup{
linkbordercolor={red}
}

\hypersetup{
    bookmarks=false,         
    unicode=false,          
    pdftoolbar=true,        
    pdfmenubar=true,        
    pdffitwindow=false,     
    pdfstartview={FitH},    
    pdftitle={Duality between Galilei and Carroll from Brane Point of View},    
    pdfauthor={Eric Bergshoeff, Luca Romano},     
    pdfsubject={},   
    pdfnewwindow=true,      
    colorlinks=false,       
    linkcolor=red,          
    citecolor=red,        
    filecolor=red,      
    urlcolor=red           
    linkbordercolor={red},
    citebordercolor={red},
    urlbordercolor={red}
}

\makeatletter

\newcommand{\Rmnum}[1]{\expandafter\@slowromancap\romannumeral #1@}
\makeatother

\theoremstyle{definition}

\theoremstyle{remark}

\theoremstyle{proposition}

\usepackage{alphalph}

\makeatletter
\newalphalph{\aalphalph}[mult]{\alphalph@alph}{26}
\newcommand{\alphalphval}[1]{%
  \@ifundefined{c@#1}{
    \aalphalph{#1}
  }{%
    \aalphalph{\value{#1}}
  }
}
\makeatother

\AtBeginEnvironment{subequations}{%
  \let\alph\alphalphval%
}

\usepackage{color}

\def\chapterautorefname~#1\null{Chap.~(#1)\null}
\def\sectionautorefname~#1\null{Sec.~(#1)\null}
\def\subsectionautorefname~#1\null{sub--Sec.~(#1)\null}
\def\figureautorefname~#1\null{Fig.~(#1)\null}
\def\tableautorefname~#1\null{Tab.~(#1)\null}
\def\equationautorefname~#1\null{eq.~(#1)\null}

\def\equationautorefname~#1\null{eq.~(#1)\null}

\NewEnviron{multieq}[1][2]{

\begin{multicols}{#1}
\begin{subequations}
\setlength{\abovedisplayskip}{-12pt}
\allowdisplaybreaks
\begin{align}
\BODY
\end{align}
\end{subequations}
\end{multicols}
\setlength{\parindent}{0pt}
}

\NewEnviron{multieqsep}[1][2]{

\begin{multicols}{#1}
\setlength{\columnseprule}{0.4pt}
\begin{subequations}
\setlength{\abovedisplayskip}{-12pt}
\allowdisplaybreaks
\begin{align}
\BODY
\end{align}
\end{subequations}
\end{multicols}
\setlength{\parindent}{0pt}
}

\DeclareMathAlphabet\mathbfcal{OMS}{cmsy}{b}{n}

\usepackage[yyyymmdd]{datetime}

\title{\bf Carroll versus Galilei from a Brane Perspective}
\date{}

\begin{document}

{\let\newpage\relax\maketitle}
\maketitle
\def\equationautorefname~#1\null{eq.~(#1)\null}
\def\tableautorefname~#1\null{tab.~(#1)\null}

\vspace{0.8cm}

\begin{center}


\renewcommand{\thefootnote}{\alph{footnote}}
{\sl\large Eric Bergshoeff$^{~1}$}\footnote{Email: {\tt e.a.bergshoeff[at]rug.nl}},
{\sl\large Jos\'e Manuel Izquierdo$^{~2}$}\footnote{Email: {\tt izquierd[at]fta.uva.es}} and
{\sl\large Luca Romano$^{~1,3}$}\footnote{Email: {\tt lucaromano2607[at]gmail.com}; address after 01-01-20: Van Swinderen Institute, Groningen University.}

\setcounter{footnote}{0}
\renewcommand{\thefootnote}{\arabic{footnote}}

\vspace{0.5cm}

${}^1${\it Van Swinderen Institute, University of Groningen\\
Nijenborgh 4, 9747 AG Groningen, The Netherlands}\\
\vskip .2truecm

${}^{2}${\it  Departamento de F\'{\i}sica Te\'orica, Universidad de Valladolid,\\
  E-47011-Valladolid, Spain}

\vskip .2truecm
${}^3${\it Instituto de F\'{\i}sica Te\'orica UAM/CSIC\\
C/ Nicol\'as Cabrera, 13--15,  C.U.~Cantoblanco, E-28049 Madrid, Spain}\\

\vspace{1.8cm}


{\bf Abstract}
\end{center}
\begin{quotation}
{\small
We show that our previous work on Galilei and Carroll gravity, apt for particles, can be generalized to   Galilei and Carroll gravity theories adapted to $p$-branes ($p=0,1,2, \cdots$). Within this wider brane perspective, we make use of a formal map, given in the literature,  between the corresponding $p$-brane Carroll and Galilei algebras where the index describing the directions longitudinal (transverse) to the Galilei brane is interchanged with the index covering the directions transverse (longitudinal) to the Carroll brane with the understanding  that the time coordinate is always among the longitudinal directions. This leads among other things in 3D to a map between Galilei  particles and Carroll strings and in 4D to a similar map between Galilei strings and Carroll strings. We show that this  formal map extends to the corresponding Lie algebra expansion of the Poincar\'e algebra and, therefore, to several extensions of the Carroll and Galilei algebras  including central extensions. We use this formal map to construct several new examples of Carroll gravity actions. Furthermore, we  discuss the symmetry between Carroll and Galilei at the level of the $p$-brane sigma model action and apply  this formal symmetry to give several examples of $3D$ and $4D$ particles and strings in  a curved Carroll background.
}
\end{quotation}

\newpage

\tableofcontents

\section{Introduction}

It is well known that the anti-de Sitter algebra allows several contractions to non-relativistic kinematical symmetry algebras \cite{Bacry:1968zf}. For zero cosmological constant, among these algebras are the Galilei and Carroll algebras. Both algebras can be obtained as a contraction of the Poincar\'e algebra which by itself is a contraction of the anti-de Sitter algebra. Whereas the Galilei algebra is obtained as a non-relativistic, i.e.~$c\rightarrow\infty$, limit of the Poincar\'e algebra, the Carroll algebra is obtained as the opposite ultra-relativistic, i.e.~$c\rightarrow 0$, limit of the same Poincar\' e algebra.

For obvious reasons the Galilei algebra has received much more attention than the Carroll algebra. However, the Carroll algebra has regained interest over the past few years. Carroll symmetries occur in a variety of special situations such as in the strong coupling limit of General Relativity \cite{Henneaux} and near spacetime singularities \cite{Damour:2002et}. They have also emerged in a recent study of the isometry group of plane gravitational waves \cite{Duval:2017els} and in a study of warped AdS holography \cite{Hofman:2014loa}.
The same Carroll symmetries play a prominent role in flat space holography \cite{Banks:2016taq}. According to \cite{Duval:2014uva} a conformal extension of the Carroll symmetries is related to the  BMS symmetries \cite{Bondi:1962px}. These BMS symmetries occur as a subgroup of the symmetries of a massless particle moving in a Carroll geometry \cite{Bergshoeff:2014jla}. Taking a special limit of general relativity, an  ultra-relativistic version of gravity, called Carroll gravity, has been constructed \cite{Bergshoeff:2017btm}. The corresponding action is related, and perhaps equal, to the one considered in \cite{Henneaux}. An alternative version of Carroll gravity has been constructed in \cite{Hartong:2015xda}. For other recent papers on Carroll symmetries, see, e.g., \cite{Bagchi:2019xfx,Roychowdhury:2019aoi,Ravera:2019ize,Kluson:2017fam}.

The Galilei and Carroll algebra look rather different from a particle point of view where time plays a special role. On the one hand, the $D$-dimensional Galilei algebra  is given by
\vskip .3truecm

\centerline{\bf The $\pmb{D}$-dimensional Galilei algebra}
\vskip .3truecm

\begin{multicols}{2}
\begin{subequations}\label{p=0Galilei}
\setlength{\abovedisplayskip}{-11pt}
\allowdisplaybreaks
\begin{align}
[J_{ab},J_{cd}]&=4\eta_{[a[c}J_{d]b]}\,,\\[.1truecm]
[J_{ab},G_{d}]&=2\eta_{d[b|}G_{|a]}\,,\\[.1truecm]
[G_{a},G_{b}]&=0\,,\\[.1truecm]
[J_{ab},P_{d}]&=2\eta_{d[b}P_{a]}\,,\\[.1truecm]
[G_{a},H]&=P_{a}\,,\\[.1truecm]
[G_{a},P_{b}]&=0\,,
\end{align}
\end{subequations}
\end{multicols}
\noindent where $\{H\,,P_a\,,J_{ab}\,,G_a\}$ are the generators  corresponding to time translations, spatial translations, spatial rotations and Galilean boosts, respectively. On the other hand, the non-zero commutation relations of the $D$-dimensional Carroll algebra are given by
\vskip .3truecm

\centerline{\bf The $\pmb{D}$-dimensional Carroll algebra}
\vskip .3truecm

\begin{multicols}{2}
\begin{subequations}\label{p=0Carroll}
\setlength{\abovedisplayskip}{-11pt}
\allowdisplaybreaks
\begin{align}
[J_{ab},J_{cd}]&=4\eta_{[a[c}J_{d]b]}\,,\\[.1truecm]
[J_{ab},G_{d}]&=2\eta_{d[b|}G_{|a]}\,,\\[.1truecm]
[G_{a},G_{b}]&=0\,,\\[.1truecm]
[J_{ab},P_{d}]&=2\eta_{d[b}P_{a]}\,,\\[.1truecm]
[G_{a},P_b]&=-\delta_{ab}H\,,\\[.1truecm]
[G_{a},H]&=-P_a\,.
\end{align}
\end{subequations}
\end{multicols}
\noindent We see that the difference between the Galilei and Carroll algebra stems from  the $G_a,P_a,H$ commutators involving the
boost generators $G_a$, the time translation generator $H$ and the spatial translation generators $P_a$. Nevertheless, within this particle perspective, an interesting
duality relation between Carroll and Galilei  has been discovered using two dual non-Einsteinian concepts of time
\cite{Duval:2014uoa}.

It has been pointed out that, although
the  Galilei and Carroll algebras look  different, the two algebras are  very similar when we give up the particle perspective and consider the two algebras from a wider  brane perspective where we do not distinguish between a time direction and the other transverse directions \cite{Barducci:2018wuj}. Instead, we decompose a general flat spacetime index into the directions longitudinal to the brane, including the time direction, and into the directions transverse to the brane. We should stress that the resulting  similarity between Carroll and Galilei that we will discuss in this work is a formal one and should be distinguished from the more physical duality advocated in \cite{Duval:2014uoa}. In this work  we will use this formal map between Galilei and Carroll to construct several new examples of Carroll gravity theories for branes.
Note that we use here the word `branes' in a kinematical way without actually writing down a sigma model action for such a brane. In the second part of this  work, we will extend the discussion to the $p$-brane sigma model action and construct new examples of sigma models describing particles and strings  in a curved Carroll background. For earlier work  on Carroll particles, strings and branes, see \cite{Bergshoeff:2014jla,Bergshoeff:2015wma,Cardona:2016ytk,Barducci:2018wuj}.

The organization of this work is as follows. In section 2, we will consider the general $p$-brane contraction of the Poincar\'e algebra where the longitudinal components of the generators scale differently with the contraction parameter than the transverse components. This automatically leads to contracted algebras with a manifest Lorentz (spatial rotational) symmetry in the longitudinal (transverse)  directions.
 Next, we review the formal map between the Carroll and Galilei algebras from this wider brane perspective \cite{Barducci:2018wuj}. We will point out that this  map extends to the Lie algebra expansion  \cite{Hatsuda:2001pp, deAzcarraga:2002xi} of the Poincar\'e algebra and therefore extends to the corresponding {\sl extended} Galilei and  Carroll algebras including central extensions.\,\footnote{\EAB{In the case of AdS this relation between Galilean and Carroll expansions was studied in \cite{Gomis:2019nih}.}}\,\footnote{Our nomenclature of the different algebras and gravity theories occurring in this work is explained in Appendix A.} In section 3, we will use this  formal map to construct several actions for (extended) Carroll gravity by using our earlier results for (extended) Galilei gravity obtained in \cite{Bergshoeff:2017btm,Bergshoeff:2019ctr}. Next, in section 4, we will discuss in which sense this symmetry between Carroll and Galilei extends at the level of the $p$-brane sigma model actions. We will give several examples of $3D$ and $4D$ particles and strings in a curved Carroll background. Finally, in section 5 we will give our conclusions.

\section{Carroll versus Galilei and the Lie Algebra Expansion}

In this section we show  the details of the  formal map between the Carroll and Galilei algebras mentioned in the introduction and point out how this formal map   extends to the corresponding Lie algebra expansion of the Poincar\'e algebra.

Our starting point is the $D$-dimensional Poincar\'e algebra with the following commutation relations:
\begin{subequations}\label{Poincare}
\begin{align}
[P_{\tilde{A}},P_{\tilde{B}}]&=0\,,\\[.1truecm]
[J_{\tilde{A}\tilde{B}},P_{\tilde{C}}]&=\eta_{\tilde{C}\tilde{B}}P_{\tilde{A}}-\eta_{\tilde{C}\tilde{A}}P_{\tilde{B}}\,,\\[.1truecm]
[J_{\tilde{A}\tilde{B}},J_{\tilde{C}\tilde{D}}]&=
4\eta_{[\tilde{A}[\tilde{C}}J_{\tilde{D}]\tilde{B}]},
\end{align}\label{eq:PoincareAlgebraCommutators}
\end{subequations}
where $P_{\tilde{A}}$ and $J_{\tilde{A}\tilde{B}}$ are the generators of spacetime translations and Lorentz transformations, respectively. The flat tilde  indices run over $\tilde{A}=0,1,\dots,D-1$.
In the case of a $p$-brane we decompose the index $\tilde A$ into indices  $A,a$ describing the directions longitudinal and transverse to the $p$-brane as follows:
 \begin{equation}
 \tilde{A}=\{A,a\}\hskip .5truecm  \textrm{with}\  A=0,1,\dots,p\  \textrm{and}\  a=p+1,\dots,D-1\,.
  \end{equation}{
  This induces the following  decomposition of the generators:
\begin{align}
J_{\tilde{A}\tilde{B}}&\rightarrow\{J_{AB}, G_{Aa},J_{ab}\}\,, \hskip 1.5truecm
 P_{\tilde{A}}\rightarrow\{H_{A}, P_{a}\}\,.
\end{align}
In terms of the decomposed generators, the non-zero commutators of the Poincar\'e algebra are given by
\begin{multicols}{2}
\begin{subequations}\label{decomposed}
\setlength{\abovedisplayskip}{-11pt}
\allowdisplaybreaks
\begin{align}
[J_{AB},J_{CD}]&=4\eta_{[A[C}J_{D]B]}\,,\\[.1truecm]
[J_{ab},J_{cd}]&=4\eta_{[a[c}J_{d]b]}\,,\\[.1truecm]
[J_{AB},G_{Cd}]&=2\eta_{C[B}G_{A]d}\,,\\[.1truecm]
[J_{ab},G_{Cd}]&=2\eta_{d[b|}G_{C|a]}\,,\\[.1truecm]
[G_{Aa},G_{Bb}]&=-\eta_{AB}J_{ab}-\eta_{ab}J_{AB}\,.\\[.1truecm]
[J_{AB},H_{C}]&=2\eta_{C[B}H_{A]}\,,\\[.1truecm]
[J_{ab},P_{d}]&=2\eta_{d[b}P_{a]}\,,\\[.1truecm]
[G_{Aa},H_{B}]&=-\eta_{AB}P_{a}\,,\\[.1truecm]
[G_{Aa},P_{b}]&=\eta_{ab}H_{A}\,,
\end{align}
\end{subequations}
\end{multicols}

Both the Galilei $p$-brane algebra and Carroll $(D-p-2)$-brane algebras are obtained through a contraction of the Poincar\'e algebra induced by the decomposition
\begin{align}\label{decomposition1}
&\mathfrak{g}=V_{0}\oplus V_{1}\,,
\end{align}
where the subspaces $V_0$ and $V_1$ are spanned by the following generators:


\begin{multicols}{2}
\setlength{\abovedisplayskip}{-15pt}
\allowdisplaybreaks
\centering
{\bf $\pmb{p}$-brane Galilei}\\
[0.5cm]
\begin{subequations}\label{decomposition2}
\begin{align}
&V_{0}=\{J_{AB},J_{ab},H_{A}\}\,,\\[.1truecm]
&V_{1}=\{G_{Ab},P_{a}\}\,,
\end{align}
\end{subequations}
\centering
{\bf ($\pmb{D-p-2}$)-brane Carroll}\\
[0.5cm]
\begin{subequations}\label{decomposition3}
\begin{align}
&V_{0}=\{J_{AB},J_{ab},P_{a}\}\,,\\[.1truecm]
&V_{1}=\{G_{Ab},H_{A}\}\,.
\end{align}
\end{subequations}
\end{multicols}
In both cases $V_0$ is a subalgebra and the commutation relations define the following  symmetric space structure:
\begin{equation}
[V_0,V_0]\subseteq V_{0}\,,\hskip 2truecm  [V_1,V_0]\subseteq V_{1}\,,\hskip 2truecm [V_1,V_1]\subseteq V_{0}\,.
\end{equation}
The contraction is now defined by first re-scaling all generators corresponding to  $V_1$ such that they are linear in a contraction parameter $\omega$ and next taking the limit $\omega\rightarrow\infty$.
For  $p=0$ and $p=D-2$, this contraction reproduces the particle or 0-brane Galilei and Carroll algebras given in eqs.~\eqref{p=0Galilei} and \eqref{p=0Carroll}, respectively.

From the above formulae it is clear that there is a formal map between the  $p$-brane Galilei contraction and  the  $(D-p-2)$-brane Carroll contraction given by the exchange of the longitudinal and transverse indices:
\begin{equation}
 A\leftrightarrow a\,,
 \end{equation}
where it is understood that $H_A\ \leftrightarrow \ P_a$. This defines the following  map between the  $p$-brane Galilei and $(D-p-2)$-brane Carroll algebras \cite{Barducci:2018wuj}:
 \begin{table}[h!]
\centering
\begin{tabular}{ccc}\label{formal}
{\bf $\pmb{p}$-brane Galilei}&$\xLeftrightarrow{\text{$\qquad A\leftrightarrow a\qquad $}}$& {\bf ($\pmb{D-p-2}$)-brane Carroll}
\end{tabular}
\end{table}

As a simple example of how this formal map explicitly works, we consider the $3D$ Carroll and Galilei algebras for particles and strings. The non-zero commutation relations are given in Table 1.

\begin{table}[!ht]
\renewcommand{\arraystretch}{1.5}
\begin{center}
\begin{tabular}{|c|c|c|c|}
\hline
\multicolumn{4}{|c|}{\bf $\pmb{D=3}$ Non-Zero Commutation Relations}\\
\hline
\multicolumn{2}{|c|}{\bf Galilei}&\multicolumn{2}{|c|}{\bf Carroll}\\
\hline
{$\pmb{p=0}$}&{$\pmb{p=1}$}&{$\pmb{p=0}$}&{$\pmb{p=1}$}\\
\hline
$[J,G_{a}]=-\epsilon_{a}^{\ b}G_{b}$&$[M,G_{A}]=\epsilon_{A}^{\ B}G_{B}$&$[J,G_{a}]=-\epsilon_{a}^{\ b}G_{b}$&$[M,G_{A}]=\epsilon_{A}^{\ B}G_{B}$\\
$[J,P_{a}]=-\epsilon_{a}^{\ b}P_{b}$&
$[M,H_{A}]=\epsilon_{A}^{\ B}H_{B}$&$[J,P_{a}]=-\epsilon_{a}^{\ b}P_{b}$&$[M,H_{A}]=\epsilon_{A}^{\ B}H_{B}$\\
$[G_{a},H]=P_{a}$&$[G_{A},H_{B}]=-\eta_{AB}P$&$[G_{a},P_{b}]=\eta_{ab}H$&$[G_{A},P]=H_{A}$\\ \hline
\end{tabular}
\end{center}
\caption{This table compares the $3D$  Carroll and Galilei algebras for particles and strings.}\label{tab:GalileiCarrollpbranecommutations}
\end{table}
From the table we conclude that there is a formal map between the $3D$ $p$-brane Galilei and $(1-p)$-brane Carroll algebra given by
\begin{subequations}
\begin{align}
J&\leftrightarrow -M\,,\hskip 1.5truecm
P\leftrightarrow H\,, \hskip 1.5truecm
H_{A}\leftrightarrow P_{a}\,,\hskip 1.5truecm
G_{A}\leftrightarrow G_{a}\,.
\end{align}
\end{subequations}
which  simply corresponds to an interchange between the longitudinal index $A$ with the transverse index $a$.

Most importantly,  since the Lie algebra expansion procedure is fully determined by the initial decomposition defined by eqs.~\eqref{decomposition1}-\eqref{decomposition3}, we conclude that  this formal map between the Carroll and Galilei algebras will continue to hold for the expanded Poincar\'e algebra, order by order, as well. We will make use of this observation to construct several new examples of (extended) Carroll gravity actions in the next section by making use of our earlier results on (extended) Galilei gravity
\cite{Bergshoeff:2017btm,Bergshoeff:2019ctr}.

\section{Examples of Carroll Gravity}
Starting from the Einstein-Hilbert action and making use of the Lie algebra expansion of the Poincar\'e algebra,  in our previous work we have constructed several examples of non-relativistic gravity actions in $3D$ and $4D$ based upon the extended   particle ($p=0$)  and string ($p=1$) Galilei algebra
\cite{Bergshoeff:2019ctr}. Below we will show how, by making use of the formal map between Carroll and Galilei, this automatically leads to similar results for ultra-relativistic Carroll gravity actions.

In the Lie algebra expansions that we  perform below we  associate the following  gauge fields  to the decomposed generators:
\begin{multicols}{2}
\begin{subequations}
\setlength{\abovedisplayskip}{-15pt}
\allowdisplaybreaks
\begin{align}
J_{AB}&\rightarrow\ \Omega_{\mu}^{AB}\,,\\[.1truecm]
J_{ab}&\rightarrow\ \Omega_{\mu}^{ab}\,,\\[.1truecm]
G_{Ab}&\rightarrow\ \Omega_{\mu}^{Ab}\,.\\[.1truecm]
P_{a}&\rightarrow\ E_{\mu}^{a}\,,\\[.1truecm]
H_{A}&\rightarrow\ \tau_{\mu}^{A}\,,
\end{align}
\end{subequations}
\end{multicols}

\subsection[$3D$ Carroll Gravity]{$\pmb{3D}$ Carroll Gravity}
Using form notation, the $3D$  Einstein-Hilbert Lagrangian is given by
\begin{equation}\label{3DEH}
\mathcal{L}_{\mathrm{EH}}=\epsilon_{\tilde{A}\tilde{B}\tilde{C}}\ R^{\tilde{A}\tilde{B}}(J)\wedge E^{\tilde{C}}\,,
\end{equation}
where  $R^{\tilde{A}\tilde{B}}(J)$ is the 2-form Lorentz curvature of the Poincar\'e algebra \eqref{Poincare} and $E^{\tilde A}$ is the 1-form Vielbein gauge field associated with the spacetime translation generators $P_{\tilde A}$.
After a particle $(p=0)$ and string $(p=1)$ decomposition we obtain, before expanding,  the following Lagrangians:
\begin{align}\label{above}
\mathcal{L}_{p=0}&=\epsilon_{ab}\left[R^{ab}(J)\wedge \tau -  2R^{a}(G)\wedge E^{b}\right]\,,\\[.1truecm]
\mathcal{L}_{p=1}&=\epsilon_{AB}\left[R^{AB}(J)\wedge E - 2R^{A}(G)\wedge \tau^{B}\right]\,,
\end{align}
where $a=1,2$ are the two directions transverse to the particle and $A=0,1$ are the two directions longitudinal to the string.

Using the notation of \cite{Bergshoeff:2019ctr},  we now expand both Lagrangians  in eq.~\eqref{above}  to lowest order corresponding to the  Carroll or Galilei algebra $\mathfrak{g}(0,1)$.\,\footnote{We use here the notation of \cite{Bergshoeff:2019ctr} where $\mathfrak{g}(p,q)$ ($p$ even and $q$ odd) is the Lie algebra that corresponds to a Lie algebra expansion where the generators of the $V_0\, (V_1)$ subspace are expanded up to powers $\omega^p\, (\omega^q)$. The case $p=0, q=1$ corresponds to the usual Wigner-In\"on\"u contraction.} This leads to the  Lagrangians given in Table 2.
\begin{table}[!ht]
\renewcommand{\arraystretch}{3}
\begin{center}
\begin{tabular}{|c|c|c|}
\hline
$\mathbfcal{L}_{\pmb p}$&\bf Galilei&\bf Carroll\\
\hline
{$\pmb{p=0}$}&$\epsilon_{ab}\ \accentset{(0)}{R}^{ab}(J)\wedge \accentset{(0)}{\tau} $&  $\epsilon_{ab}\left[\accentset{(0)}{R}^{ab}(J)\wedge \accentset{(1)}{\tau} -  2\accentset{(1)}{R}^{a}(G)\wedge \accentset{(0)}{E}^{b}\right]$ \\
{$\pmb{p=1}$}&$\epsilon_{AB}\left[\accentset{(0)}{R}^{AB}(J)\wedge \accentset{(1)}{E} - 2\accentset{(1)}{R}^{A}(G)\wedge \accentset{(0)}{\tau}^{B}\right]$&$\epsilon_{AB}\  \accentset{(0)}{R}^{AB}(J)\wedge \accentset{(0)}{E}$\\
\hline

\end{tabular}
\end{center}
\caption[]{This table gives the Lagrangians resulting from  the lowest order Lie algebra expansion of the $3D$ Einstein-Hilbert Lagrangian \eqref{3DEH} corresponding to the $3D$ Carroll and Galilei algebra for both particles $(p=0)$  and strings $(p=1)$. }\label{tab:GalileiCarrollActionsD=3}
\end{table}

\noindent From Table 2 we deduce that, would we have restricted ourselves to particles only, the results for the Galilei and Carroll gravity Lagrangians are a-symmetric \cite{Bergshoeff:2019ctr}. From the first line in the above table, we read that the Galilei gravity Lagrangian contains only one term whereas the Carroll gravity Lagrangian contains two terms. However, incorporating also strings, we see that the symmetry is restored provided we connect the particle Galilei Lagrangian in the first line of the table to the string Carroll Lagrangian in the second line. Both for Carroll and Galilei we have one Lagrangian with one term and a second Lagrangian with two terms.

\subsection[$3D$ Extended Carroll Gravity]{$\pmb{3D}$ Extended Carroll Gravity}

It is straightforward to go beyond the lowest order in the Lie algebra expansion of the $3D$  Einstein-Hilbert  Lagrangian \eqref{3DEH}.\footnote{This subsection has some overlap with the  recent paper \cite{Gomis:2019nih}.} In particular, in order for these actions to be invariant under the corresponding extended algebras we should consider the truncation to $\mathfrak{g}(2,1)$ for the extended $p=0$ Galilei and $p=1$ Carroll Lagrangians and to $\mathfrak{g}(2,3)$ for the extended $p=1$ Galilei  and extended $p=0$ Carroll Lagrangians. The explicit expressions for these Lagrangians are given in Table 3.

\begin{table}[!ht]
\renewcommand{\arraystretch}{3}
\begin{center}
\resizebox{\textwidth}{!}{
\begin{tabular}{|c|c|c|}
\hline
$\mathbfcal{L}_{\pmb p}$&\bf Galilei&\bf Carroll\\
\hline
{$\pmb{p=0}$}&$\epsilon_{ab}\bigg[\accentset{(2)}{R}^{ab}(J)\wedge \accentset{(0)}{\tau}+\accentset{(0)}{R}^{ab}(J)\wedge \accentset{(2)}{\tau}+$&
$\epsilon_{ab}\bigg[\accentset{(0)}{R}^{ab}(J)\wedge \accentset{(3)}{\tau}+\accentset{(2)}{R}^{ab}(J)\wedge \accentset{(1)}{\tau} +$\\

&$ -  2\accentset{(1)}{R}^{a}(G)\wedge \accentset{(1)}{E}^{b}\bigg]$&$-  2\accentset{(1)}{R}^{a}(G)\wedge \accentset{(2)}{E}^{b} -  2\accentset{(3)}{R}^{a}(G)\wedge \accentset{(0)}{E}^{b}\bigg]$\\
{ $\pmb{p=1}$}&
$\epsilon_{AB}\bigg[\accentset{(0)}{R}^{AB}(J)\wedge \accentset{(3)}{E}+\accentset{(2)}{R}^{AB}(J)\wedge \accentset{(1)}{E}+$&$\epsilon_{AB}\bigg[\accentset{(2)}{R}^{AB}(J)\wedge \accentset{(0)}{E}+\accentset{(0)}{R}^{AB}(J)\wedge \accentset{(2)}{E}+$\\
&$- 2\accentset{(1)}{R}^{A}(G)\wedge \accentset{(2)}{\tau}^{B}-2\accentset{(3)}{R}^{A}(G)\wedge \accentset{(0)}{\tau}^{B}
\bigg]$&$ - 2\accentset{(1)}{R}^{A}(G)\wedge \accentset{(1)}{\tau}^{B}\bigg]$\\
\hline
\end{tabular}
}
\end{center}
\caption[]{This table gives the Lagrangians resulting from   the next to lowest order Lie algebra expansion of the $3D$ Einstein-Hilbert Lagrangian \eqref{3DEH} corresponding to the $3D$ extended Carroll and Galilei algebra for both particles $(p=0)$ and strings $(p=1)$. }\label{tab:GalileiCarrollActionsD=3NextOrder}
\end{table}

The particle ($p=0$) Lagrangian for extended Galilei gravity given in the first line of Table 3 is, using our old nomenclature,  the extended Bargmann Lagrangian  constructed in
\cite{Papageorgiou:2009zc,Bergshoeff:2016lwr,Hartong:2016yrf}. Using our formal map, it immediately leads to the corresponding result
for the extended string ($p=1$) Carroll gravity Lagrangian given in the second line of the  table.

\subsection[$4D$ Carroll Gravity]{$\pmb{4D}$ Carroll Gravity}
Our starting point is the $4D$ Einstein-Hilbert Lagrangian
\begin{equation}\label{4DEH}
\mathcal{L}_{\mathrm{EH}}=\epsilon_{\tilde{A}\tilde{B}\tilde{C}\tilde{D}}\ R^{\tilde{A}\tilde{B}}(J)\wedge E^{\tilde{C}}\wedge E^{\tilde{C}}\,.
\end{equation}
After decomposing the fields, before expanding, we obtain the following Lagrangians for particles $(p=0)$, strings $(p=1)$ and membranes $(p=2)$:
\begin{subequations}
\begin{align}
\mathcal{L}_{p=0}=&\,\epsilon_{abc}\left[ -R^{ab}(J)\wedge E^{c}\wedge \tau +R^{a}(G)\wedge E^{b}\wedge E^{c}\right]\,,\\[.1truecm]
\mathcal{L}_{p=1}=&\,\epsilon_{ABab}\left[ R^{ab}(J)\wedge \tau^{A}\wedge \tau^{B} +
R^{AB}(J)\wedge E^{a}\wedge E^{b}-4R^{Aa}(G)\wedge \tau^{B}\wedge E^{b}
\right]\,,\\[.1truecm]
\mathcal{L}_{p=2}=&-\epsilon_{ABC}\left[R^{AB}(J)\wedge \tau^{C}\wedge E +R^{A}(G)\wedge \tau^{B}\wedge \tau^{C}\right]\,.
\end{align}
\end{subequations}
After expanding we find, in lowest order, for the  Carroll and Galilei algebras the results given in Table 4.

\begin{table}[!ht]
\renewcommand{\arraystretch}{3}
\begin{center}
\resizebox{\textwidth}{!}{
\begin{tabular}{|c|c|c|}
\hline
$\mathbfcal{L}_{\pmb p}$&\bf Galilei&\bf Carroll\\
\hline
{$\pmb{p=0}$}&$-\epsilon_{abc}\ \accentset{(0)}{R}^{ab}(J)\wedge \accentset{(1)}{E}^{c}\wedge \accentset{(0)}{\tau}$&$\epsilon_{abc}\left[ -\accentset{(0)}{R}^{ab}(J)\wedge \accentset{(0)}{E}^{c}\wedge \accentset{(1)}{\tau} +\accentset{(1)}{R}^{a}(G)\wedge \accentset{(0)}{E}^{b}\wedge \accentset{(0)}{E}^{c}\right]$\\
{$\pmb{p=1}$}&$\epsilon_{ABab}\ \accentset{(0)}{R}^{ab}(J)\wedge  \accentset{(0)}{\tau}^{A}\wedge  \accentset{(0)}{\tau}^{B}$&
$\epsilon_{ABab}\ \accentset{(0)}{R}^{AB}(J)\wedge  \accentset{(0)}{E}^{a}\wedge  \accentset{(0)}{E}^{b}$\\
{$\pmb{p=2}$}&$-\epsilon_{ABC}\left[\accentset{(0)}{R}^{AB}(J)\wedge \accentset{(0)}{\tau}^{C}\wedge \accentset{(1)}{E} +\accentset{(1)}{R}^{A}(G)\wedge \accentset{(0)}{\tau}^{B}\wedge \accentset{(0)}{\tau}^{C}\right]$&$-\epsilon_{ABC}\ \accentset{(0)}{R}^{AB}(J)\wedge \accentset{(1)}{\tau}^{C}\wedge \accentset{(0)}{E}$\\
\hline
\end{tabular}
}
\end{center}
\caption[]{This table gives the Lagrangians resulting from  the lowest order Lie algebra expansion of the $4D$ Einstein-Hilbert Lagrangian \eqref{4DEH}  corresponding to the $4D$  Carroll and Galilei algebra for both particles $(p=0)$, strings $(p=1)$ and membranes $(p=2)$.}\label{tab:GalileiCarrollActionsD=4}
\end{table}

Like in $3D$, we find that, after including all branes and not only particles, the results for the Carroll and Galilei gravity Lagrangians  are  symmetric. Using our formal map, the particle (membrane) Galilei gravity Lagrangian given in the left column of Table 4 is mapped to the membrane (particle) Carroll Lagrangian given in the right column of the table whereas the string Galilei and Carroll Lagrangians are mapped onto each other.

\subsection[$4D$ Extended Carroll Gravity]{$\pmb{4D}$ Extended Carroll Gravity}
In order for the next to lowest order action to be invariant we should consider the $\mathfrak{g}(2,3)$ algebra for particles ($p=0)$ and membranes $(p=2)$  and the $\mathfrak{g}(2,1)$ algebra for strings $(p=1)$ both for Carroll and Galilei. The result for the different  Lagrangians that follow from the Lie algebra expansion are given in Table 5.

\begin{table}[!ht]
\renewcommand{\arraystretch}{3.5}
\begin{center}
\resizebox{\textwidth}{!}{
\begin{tabular}{|c|c|c|}
\hline
$\mathbfcal{L}_{\pmb p}$&\bf Galilei&\bf Carroll\\
\hline
{$\pmb{p=0}$}&$\epsilon_{abc}\bigg[-\accentset{(0)}{R}^{ab}(J)\wedge \accentset{(3)}{E}^{c}\wedge \accentset{(0)}{\tau} -\accentset{(2)}{R}^{ab}(J)\wedge \accentset{(1)}{E}^{c}\wedge \accentset{(0)}{\tau}+$&$\epsilon_{abc}\bigg[-\accentset{(0)}{R}^{ab}(J)\wedge \accentset{(0)}{E}^{c}\wedge \accentset{(3)}{\tau} -\accentset{(2)}{R}^{ab}(J)\wedge \accentset{(0)}{E}^{c}\wedge \accentset{(1)}{\tau}+$\\
&$-\accentset{(0)}{R}^{ab}(J)\wedge \accentset{(1)}{E}^{c}\wedge \accentset{(2)}{\tau} +\accentset{(1)}{R}^{a}(G)\wedge \accentset{(1)}{E}^{b}\wedge \accentset{(1)}{E}^{c}\bigg]$ &$-\accentset{(0)}{R}^{ab}(J)\wedge \accentset{(2)}{E}^{c}\wedge \accentset{(1)}{\tau}+2\accentset{(1)}{R}^{a}(G)\wedge \accentset{(2)}{E}^{b}\wedge \accentset{(0)}{E}^{c}+$\\
&&$+\accentset{(3)}{R}^{a}(G)\wedge \accentset{(0)}{E}^{b}\wedge \accentset{(0)}{E}^{c}\bigg]$\\
{$\pmb{p=1}$}&
$\epsilon_{ABab}\bigg[\accentset{(2)}{R}^{ab}(J)\wedge \accentset{(0)}{\tau}^{A}\wedge \accentset{(0)}{\tau}^{B}+2\accentset{(0)}{R}^{ab}(J)\wedge \accentset{(2)}{\tau}^{A}\wedge \accentset{(0)}{\tau}^{B} +$
&$\epsilon_{ABab}\bigg[\accentset{(0)}{R}^{ab}(J)\wedge \accentset{(1)}{\tau}^{A}\wedge \accentset{(1)}{\tau}^{B} +
\accentset{(2)}{R}^{AB}(J)\wedge \accentset{(0)}{E}^{a}\wedge \accentset{(0)}{E}^{b}+$\\
&$+
\accentset{(0)}{R}^{AB}(J)\wedge \accentset{(1)}{E}^{a}\wedge \accentset{(1)}{E}^{b}-4\accentset{(1)}{R}^{Aa}(G)\wedge \accentset{(0)}{\tau}^{B}\wedge \accentset{(1)}{E}^{b}
\bigg]$&$+2\accentset{(0)}{R}^{AB}(J)\wedge \accentset{(2)}{E}^{a}\wedge \accentset{(0)}{E}^{b}-4\accentset{(1)}{R}^{Aa}(G)\wedge \accentset{(1)}{\tau}^{B}\wedge \accentset{(0)}{E}^{b}
\bigg]$\\
{$\pmb{p=2}$}&
$-\epsilon_{ABC}\bigg[\accentset{(0)}{R}^{AB}(J)\wedge \accentset{(0)}{\tau}^{C}\wedge \accentset{(3)}{E}+\accentset{(2)}{R}^{AB}(J)\wedge \accentset{(0)}{\tau}^{C}\wedge \accentset{(1)}{E}
+$&$-\epsilon_{ABC}\bigg[
\accentset{(2)}{R}^{AB}(J)\wedge \accentset{(1)}{\tau}^{C}\wedge \accentset{(0)}{E}+
\accentset{(0)}{R}^{AB}(J)\wedge \accentset{(3)}{\tau}^{C}\wedge \accentset{(0)}{E}+$\\
&$+\accentset{(0)}{R}^{AB}(J)\wedge \accentset{(2)}{\tau}^{C}\wedge \accentset{(1)}{E} +2\accentset{(1)}{R}^{A}(G)\wedge \accentset{(2)}{\tau}^{B}\wedge \accentset{(0)}{\tau}^{C}+$&$\accentset{(0)}{R}^{AB}(J)\wedge \accentset{(1)}{\tau}^{C}\wedge \accentset{(2)}{E} +\accentset{(1)}{R}^{A}(G)\wedge \accentset{(1)}{\tau}^{B}\wedge \accentset{(1)}{\tau}^{C}\bigg]$\\
&$+\accentset{(3)}{R}^{A}(G)\wedge \accentset{(0)}{\tau}^{B}\wedge \accentset{(0)}{\tau}^{C}\bigg]$&\\
\hline
\end{tabular}
}
\end{center}
\caption[]{This table gives the Lagrangians resulting from  the next to lowest order in the Lie algebra expansion  of the $4D$ Einstein-Hilbert Lagrangian \eqref{4DEH} corresponding to the $4D$ extended Carroll and Galilei algebra for both particles $(p=0)$, strings $(p=1)$ and membranes $(p=2)$.}\label{tab:GalileiCarrollActionsD=4NextOrder}
\end{table}

The  particle $(p=0)$ extended Galilei gravity Lagrangian given in Table 5 is the one constructed in \cite{Hansen:2018ofj}. We use here the first-order formulation given in \cite{Bergshoeff:2016lwr}. The string $(p=1)$ extended Galilei gravity  Lagrangian is, using the old nomenclature,  the extended string NC gravity Lagrangian constructed in \cite{Bergshoeff:2018vfn}. Table 5 shows how each extended Galilei gravity  Lagrangian leads to a corresponding invariant extended Carroll gravity  Lagrangian.

\subsection{Higher Order Expansion and Invariance Conditions}
The results that we have obtained in the previous sections, considering action terms at second or third order in the expansion parameter, could be easily generalized to study higher order terms . In particular, following the approach discussed in \cite{Bergshoeff:2019ctr}, it is possible to define a set of conditions between the action of order $n$ and the truncation of order $N_{0}, N_{1}$, such that the order $n$ action term is invariant under the algebra $\mathfrak{g}(N_{0},N_{1})$. Since the derivation follows exactly the same steps discussed in \cite{Bergshoeff:2019ctr} we refer to that work for details and we just list the results:
\begin{subequations}\label{invcond}
\begin{align}
&p=D-1&&n\leqslant N_{1}+D+1\\
&\nonumber\\
&p=D-2&&\left\{
\begin{array}{lcl}
D=3 &&n \leqslant N_{0}\\ [0.2cm]
D\neq 3 &&n\leqslant N_{1}+D-4
\end{array}\right.\\
&\nonumber\\
&1 < p\leqslant D-3 &&n\leqslant N_{1}+p-2\\
&\nonumber\\
&p=1&&n \leqslant N_{0}\\
&\nonumber\\
&p=0&&n\leqslant N_{1}\ .
\end{align}
\end{subequations}
We remark that the conditions can be obtained just exchanging $p$ with $D-p-2$ , with the exception of the $p=D-1$ which has no Galilei dual.\footnote{The case $p=D-1$ is special since it is dual to a Galilei instanton, $p=-1$ corresponding to Euclidean signature for the flat space.} From these conditions, we may also derive that, for given $(D,p)$, the
smallest algebra consistent with the invariance of the Lagrangian
density at order $n$ is given by
\begin{subequations}\label{invcond2}
\begin{align}
&p=D-1&&\mathfrak{g}(n-D, n-D+1)\\[4pt]
&p>1&& \mathfrak{g}(n-p+1, n-p+2)\\[4pt]
&p=1&&\mathfrak{g}(n,n-1)\\[4pt]
&p=0 &&\mathfrak{g}(n-1,n)\ .
\end{align}
\end{subequations}
In the case with $n=0$ it is understood that the algebra
$\mathfrak{g}(0,-1)$ corresponds to $V_{0}$.

\FloatBarrier
\section[$p$-brane Sigma Models]{$\pmb{p}$-brane Sigma Models}

We now consider the symmetry between Carroll and Galilei at the level of the $p$-brane sigma model. Our starting point is the relativistic $p$-brane Polyakov-type Lagrangian\,\footnote{We use here a unified notation including particles, i.e.~$p=0$. In that case one should take $h_{00} = e^2$ and $T=m^2$ where $e$ is the worldline Einbein and $m$ is the mass of the particle.}
\begin{equation}\label{Polyakov}
\mathcal{L}_{\textrm{Pol}} = \sqrt{h} h^{\alpha\beta}\partial_{\alpha}X^{\mu}\partial_{\beta}X^{\nu}E_{\mu}^{\tilde{A}}E_{\nu}^{\tilde{B}}\eta_{\tilde{A}\tilde{B}}
-(p-1)T\sqrt{h}\,,
\end{equation}
where $T$ is the $p$-brane tension, $X^\mu(\sigma)$ are the embedding coordinates, $E_\mu^{\tilde A}$ are the relativistic Vielbeine and $h_{\alpha\beta}$ is the worldvolume metric with $h=|\det h_{\alpha\beta}|$. Since this Lagrangian is only invariant under Lorentz rotations it is not clear how to perform a full Lie algebra expansion of this Lagrangian with respect to the Poincar\'e algebra.
Another complicating factor is that, unlike in the case of the EH Lagrangian considered in the previous section, one cannot relate
the $P$-translations of the Poincar\'e algebra to the general coordinate transformations  via so-called trivial symmetries \cite{Bergshoeff:2019ctr}. In fact, the sigma models considered in this section are strictly speaking not invariant under general coordinate transformations.  They are so-called `sigma model symmetries'. For all these reasons we will avoid in this section using the Lie algebra expansion method.

In the next two subsections  we will discuss $p$-brane sigma models first without and next with central charge symmetries using different methods. As we will see, the $p$-brane sigma models without central charge symmetries are based on the Galilei and Carroll algebras. On the other hand, the $p$-brane sigma models with central charge symmetries are based on what we will call the {\sl enhanced} Galilei and Carroll algebra.\,\footnote{See Appendix A for our nomenclature. The enhanced Galilei algebra is the well-known Bargmann algebra.} These algebras are smaller than the extended Galilei and Carrol algebras discussed in the previous section and do not have an invariant action.

\subsection{Sigma Models Without Central Charge Symmetry}

To define a non-relativistic limit of the sigma model Lagrangian \eqref{Polyakov}, we redefine the background fields using the velocity of light $c$ in the same way that  we defined the non-relativistic limit of the Einstein-Hilbert Lagrangian leading to Galilei or Carroll gravity \cite{Bergshoeff:2017btm}. Substituting this lowest-order expansion into the sigma model Lagrangian \eqref{Polyakov} we obtain the following Lagrangian:
\begin{align}\label{PL}
\mathcal{L}_{\textrm{Pol}}(\textrm{lowest}) =\sqrt{h} h^{\alpha\beta}\bigg(c^{2}\tau^{A}_{\alpha}\tau_{\beta}^{B}\eta_{AB}
+E_{\alpha}^{a}E^{b}_{\beta}\delta_{ab}\bigg)-(p-1)T\sqrt{h}\,,
\end{align}
where we have defined the pull-backs
\begin{equation}
\tau^{A}_{\alpha}\equiv\partial_{\alpha}X^{\mu}\tau_{\mu}^{A}\,,\hskip 2truecm
E^{a}_{\alpha}\equiv\partial_{\alpha}X^{\mu}E_{\mu}^{a}\,.
\end{equation}

Taking the non-relativistic limit $c\rightarrow\infty$, we  consider two cases.\,\footnote{We only consider NR limits where we preserve the relativistic symmetries of the world-sheet.}  Either, we first eliminate the $c^2$ factor in the first leading term by a rescaling of $h_{\alpha\beta}$ and $T$ ending up with a $c^{-2}$ factor in front of the second term  and then take the limit or we effectively remove the leading term by using  the following equivalent Lagrangian that contains an auxiliary field $\lambda^{A\alpha}$ (before taking the limit):\,\footnote{In the second case, we change  the sign of the $E^2$ term in the Lagrangian  before taking the limit in order to avoid, after taking the limit, getting negative signs under the square root in the Nambu-Goto formulation that follows from solving for the world-sheet metric $h_{\alpha\beta}$. We follow here \cite{Batlle:2017cfa}. }

\begin{align}
& -\sqrt{h} h^{\alpha\beta}\bigg[ E_{\alpha}^{a}E_{\beta}^{b}\delta_{ab}-\eta_{AB}\bigg(\frac{1}{c^2}\lambda^{A\alpha}\lambda^{B\beta}
-2\lambda^{A\alpha}\tau^{B\beta}\bigg)\bigg]-(p-1)T\sqrt{h}\,.
\end{align}
We next take the limit. This leads to the two cases given in the left column of Table 6. The first row corresponds to the NR `particle limit' of branes that  has been discussed in \cite{Batlle:2016iel}.\,\footnote{Note that for $p=0$ and flat Newtonian spacetime the Lagrangian is a total derivative.} The second row corresponds, for $p=0$, to the Souriau (zero spin) Galilean particle \cite{Souriau} which has  been discussed in  \cite{Batlle:2017cfa}.

Similarly, when taking the ultra-relativistic limit $c\rightarrow 0$, we either take this limit straight-away in the Polyakov Lagrangian \eqref{PL} such that we end up with the second  term or we first redefine $h_{\alpha\beta}$ and $T$ such that there is a  $c^{-2}$ factor  in front of the second term which becomes leading and constrain this leading term to be zero by introducing a Lagrange multiplier field $\lambda^{a\alpha}$ as follows:

\begin{align}
& \sqrt{h} h^{\alpha\beta}\bigg[ \tau_{\alpha}^{A}\tau_{\beta}^{B}\eta_{AB}-
\delta_{ab}\bigg(c^2 \lambda^{a\alpha}\lambda^{b\beta}-2\lambda^{a\alpha}E^{b\beta}\bigg)\bigg]-(p-1)T\sqrt{h}
\end{align}
and then take the limit. This leads to the two cases given in the right column of Table 6. The first row corresponds to a Carroll system that, for $p=0$, has been discussed in \cite{deBoer:2017ing}.\,\footnote{Private communication with S.~Vandoren.}

\begin{table}[!ht]
\renewcommand{\arraystretch}{2}
\begin{center}
\begin{tabular}{|c|c|c|}
\hline
&{\bf Galilei}&{\bf Carroll}\\
\hline
{\bf Leading Term}&$h^{\alpha\beta}\tau^{A}_{\alpha}\tau_{\beta}^{B}\eta_{AB}$&$h^{\alpha\beta}E^{a}_{\alpha}E_{\beta}^{b}\delta_{ab}$\\
\hline
 &$h^{\alpha\beta}E^{a}_{\alpha}E_{\beta}^{b}\delta_{ab}$&$h^{\alpha\beta}\tau^{A}_{\alpha}\tau_{\beta}^{B}\eta_{AB}$\\
\multirow{-2}{*}{\bf Sub-leading Term}&$\tau_{\alpha}^{A}=0$ &$E_{\alpha}^{a}=0$ \\
\hline

\end{tabular}
\end{center}
\caption{This table summarizes the four different ways in which we can define the Galilei and Carroll limits of a $p$-brane at lowest order in an expansion of $c$ consistent with the Carroll and Galilei algebras. In all cases we have not given the $(p-1)T\sqrt{h}$ term that is common to all Lagrangians.}\label{tab:}
\end{table}

We note that the four cases given in Table 6 are  consistent with the formal map between Carroll and Galilei discussed in section 2. Note also that to show that the Lagrangians in the second row of the table are invariant under Galilean or Carroll boosts one needs to use the given constraint equation. 

%

\subsection{Sigma Models With A (Non-)Central Charge Symmetry}

We now wish to consider examples of particle and string  sigma models with an enhanced Galilei or Carroll algebra of symmetries involving a central extension.\,
Although the enhanced Galilei algebra, i.e.~the Bargmann algebra,
does not occur in the Lie algebra expansion of the Poincar\'e algebra \EAB{as discussed in this paper}, it is clear, based on the symmetry
\begin{equation}
H_A \leftrightarrow P_a\,, \hskip 1.5truecm J_{AB} \leftrightarrow J_{ab}\,, \hskip 1.5truecm G_{aA} \leftrightarrow G_{aA}
\end{equation}
of the decomposed Poincar\'e algebra \eqref{decomposed}, that for every $p$-brane enhanced Galilei algebra there is a corresponding $(D-p-2)$-brane  enhanced  Carroll  algebra.
In Table 7 we have given the explicit commutations relations of the enhanced Galilei and enhanced Carroll algebras for $3D$ particles and strings. Note that these algebras have both central charge as well as  non-central charge generators.\,\footnote{We call a generator non-central charge if it has non-zero commutation relations with  the other generators due to its index structure.}
\vskip .2truecm

\begin{table}[!ht]
\renewcommand{\arraystretch}{1.4}
\begin{center}
\begin{tabular}{|c|c|c|c|}
\hline
 &&\bf Galilei&\bf Carroll\\

\cline{2-4}
&$\begin{aligned}[c]
&\pmb{D=4}\\
&\pmb{p=1}
\end{aligned}$&
$\begin{aligned}[c]
\\
[J,P_{a}]&=-\epsilon_{a}{}^{b}P_{b}\\
[J',H_{A}]&=-\epsilon_{A}{}^{B}H_{B}\\
[J',G_{Aa}]&=-\epsilon_{A}{}^{B}G_{Ba}\\
[J,G_{Aa}]&=-\epsilon_{a}{}^{b}G_{Ab}\\
[G_{Aa},H_{B}]&=-\eta_{AB}P_{a}\\
[G_{Aa},G_{Bb}]&=\delta_{ab}Z_{[AB]}\\
[J',M_{A}]&=-\epsilon_{A}{}^{B}M_{B}\\
[J',Z_{AB}]&=-\epsilon_{A}{}^{C}Z_{CB}-\epsilon_{B}{}^{C}Z_{AC}\\
[G_{Aa},P_{b}]&=\delta_{ab}M_{A}\\
[H_{A},Z_{BC}]&=2\eta_{AC}M_{B}-\eta_{BC}M_{A}\\
\\
\end{aligned}$&
$\begin{aligned}[c]
\\
[J,P_{a}]&=-\epsilon_{a}{}^{b}P_{b}\\
[J',H_{A}]&=-\epsilon_{A}{}^{B}H_{B}\\
[J',G_{Aa}]&=-\epsilon_{A}{}^{B}G_{Ba}\\
[J,G_{Aa}]&=-\epsilon_{a}{}^{b}G_{Ab}\\
[G_{Aa},P_{b}]&=-\delta_{ab}H_{A}\\
[G_{Aa},G_{Bb}]&=\eta_{AB}Z_{[ab]}\\
[J,M_{a}]&=-\epsilon_{a}{}^{b}M_{b}\\
[J,Z_{ab}]&=-\epsilon_{a}{}^{c}Z_{cb}-\epsilon_{b}{}^{c}Z_{ac}\\
[G_{Aa},H_{B}]&=\eta_{AB}M_{a}\\
[P_{a},Z_{bc}]&=2\delta_{ac}M_{b}-\delta_{bc}M_{a}\\
\\
\end{aligned}$
\\
\cline{2-4}
&$\begin{aligned}[c]
&\pmb{D=3}\\
&\pmb{p=0}
\end{aligned}$&
$\begin{aligned}[c]
\\
[J,P_{a}]&=-\epsilon_{a}{}^{b}P_{b}\\
[J,G_{a}]&=-\epsilon_{a}{}^{b}G_{b}\\
[G_{a},H]&=P_{a}\\
[G_{a},P_{b}]&=\delta_{ab}M\\
\\
\end{aligned}$&
$\begin{aligned}[c]
\\
[J,P_{a}]&=-\epsilon_{a}{}^{b}P_{b}\\
[J,G_{a}]&=-\epsilon_{a}{}^{b}G_{b}\\
[J,M_{a}]&=-\epsilon_{a}{}^{b}M_{b}\\
[J,Z_{ab}]&=-\epsilon_{a}{}^{c}Z_{cb}-\epsilon_{b}{}^{c}Z_{ac}\\
[G_{a},G_{b}]&=Z_{[ab]}\\
[G_{a},P_{b}]&=-\delta_{ab}H\\
[G_{a},H]&=M_{a}\\
[P_{a},Z_{bc}]&=2\delta_{ac}M_{b}-\delta_{bc}M_{a}\\
\\
\end{aligned}$\\
\cline{2-4}
\multirow{-4}{*}{\rotatebox[]{90}{ \bf Commutation Rules \hspace{-9cm}}}&$\begin{aligned}[c]
&\pmb{D=3}\\
&\pmb{p=1}
\end{aligned}$&
$\begin{aligned}[c]
\\
[J',H_{A}]&=-\epsilon_{A}{}^{B}H_{B}\\
[J',G_{A}]&=-\epsilon_{A}{}^{B}G_{B}\\
[J',M_{A}]&=-\epsilon_{A}{}^{B}M_{B}\\
[J',Z_{AB}]&=-\epsilon_{A}{}^{C}Z_{CB}-\epsilon_{B}{}^{C}Z_{AC}\\
[G_{A},G_{B}]&=Z_{[AB]}\\
[G_{A},H_{B}]&=-\eta_{AB}P\\
[G_{A},P]&=M_{A}\\
[H_{A},Z_{BC}]&=2\eta_{AC}M_{B}-\eta_{BC}M_{A}\\
\\
\end{aligned}$&$\begin{aligned}[c]
\\
[J',H_{A}]&=-\epsilon_{A}{}^{B}H_{B}\\
[J',G_{A}]&=-\epsilon_{A}{}^{B}G_{B}\\
[G_{A},H_{B}]&=-\eta_{AB}M\\
[G_{A},P]&=H_{A}\\
\\
\end{aligned}$\\
\hline
\end{tabular}
\end{center}
\caption[]{This table summarizes the non-zero commutation relations defining the $p=0$ and $p=1$ enhanced Galilei and enhanced Carroll algebras in three and four dimensions discussed in the text. The generators $Z_{AB}$ of the enhanced $p=1$ Galilei algebra and  $Z_{ab}$ of the $p=0$ enhanced Carroll algebra are traceless 2-tensors, i.e.~$Z^A{}_A= Z^a{}_a=0$.}\label{tab:algebras}
\end{table}

 A new feature of Carroll and Galilei $p$-brane sigma models with (non-)central charge symmetries is that, compared to the previous subsection, they contain extra terms describing the coupling of the (non-)central charge gauge field to the $p$-brane via a Wess-Zumino term. Since this Wess-Zumino term looks different for different dimensions of the worldvolume, this complicates comparing Galilei with Carroll. The only cases where the symmetry between Galilei and Carroll is restored are those for which the number of longitudinal directions is equal to the number of transverse directions, i.e.~particles in 2 dimensions, strings in 4 dimensions, membranes in 6 dimensions etc. For those branes  the formal symmetry between Carroll and Galilei sigma models  is very natural due to the fact that  under the Galilei/Carroll map branes  are mapped onto branes with the same spatial extension and therefore with a  similar Wess-Zumino term.

 As a warming up example, we first consider the Galilei and Carroll particle in 2 dimensions:

 \vskip .2truecm
 \noindent {\bf A.\ \ $\pmb{2D}$ Enhanced Galilei particle}
 \vskip .2truecm

 The Lagrangian of a $2D$ enhanced Galilei particle coupled to $2D$ enhanced Galilei gravity with field content $\{\tau_{\hat\mu}\,, E_{\hat\mu}\,, m_{\hat\mu}\}$ is given by
\begin{subequations}\label{2DGP}
\begin{align}
 {\cal L} &= m e^{-1} \dot X^\mu \dot X^\nu \bigl[ E_{\hat\mu} E_{\hat\nu} - 2 m_{(\hat\mu}\tau_{\hat\nu)}\big]\,,\\[.1truecm]
&\textrm{with}\ \   e= \dot X^\mu\tau_{\hat\mu} \hskip .5truecm \textrm{and} \hskip .5truecm\partial_{[\hat \mu}\tau_{\hat\nu]}=0\, ,
\end{align}
\end{subequations}
where we adopt the convention
\begin{align}
m_{(\hat\mu}\tau_{\hat\nu)}=\frac{1}{2}(m_{\hat\mu}\tau_{\hat\nu}+m_{\hat\nu}\tau_{\hat\mu}).
\end{align}
This Lagrangian is invariant under the following Galilean boosts and central charge transformations:
\begin{eqnarray}
\delta E_{\hat\mu} &=& \lambda \tau_{\hat \mu}\,,\\[.1truecm]
\delta m_{\hat\mu} &=& \partial_{\hat\mu}\sigma +\lambda E_{\hat\mu} \,.
\end{eqnarray}
We note that the proof of invariance works in any dimension provided we assign the transverse Vielbein  an additional transverse index.

 \vskip .2truecm
 \noindent {\bf B.\ \ $\pmb{2D}$ Enhanced Carroll particle}
 \vskip .2truecm
 Under the formal Galilei/Carroll map the $2D$ enhanced Galilei particle \eqref{2DGP} is mapped to the following $2D$ enhanced Carroll particle coupled to a $2D$ Carroll background with field content $\{\tau_{\hat\mu}\,, E_{\hat\mu}\,, n_{\hat\mu}\}$:
 \begin{subequations}\label{2DCP}
\begin{align}
 {\cal L} &= m e^{-1} \dot X^\mu \dot X^\nu \bigl[ \tau_{\hat\mu} \tau_{\hat\nu} - 2 n_{(\hat\mu}E_{\hat\nu)}\big]\,,\\[.1truecm]
 &\textrm{with}\ \  e= \dot X^\mu E_{\hat\mu}\hskip .5truecm \textrm{and} \hskip .5truecm \partial_{[\hat \mu} E_{\hat\nu]}=0\,.
\end{align}
\end{subequations}
This Carroll Lagrangian is invariant under the following Carroll boosts and central charge transformations:
\begin{eqnarray}
\delta \tau_{\hat\mu} &=& \lambda E_{\hat \mu}\,,\\[.1truecm]
\delta n_{\hat\mu} &=& \partial_{\hat\mu}\sigma +\lambda \tau_{\hat\mu} \,.
\end{eqnarray}
Unlike the enhanced Galilean particle, the proof of invariance of the enhanced Carroll particle only works in $2D$. This is due to the fact that, when going to higher dimensions,  the solution for the Einbein is inconsistent with the assignment of a transverse index to the transverse Vielbein. We will see below how an expression for the 3D Carroll particle can be derived by  the double dimensional reduction of a 4D Carroll string.

  \vskip .2truecm

 This is the end of the story for Carroll particles in $2D$ whose Lagrangian \eqref{2DCP}  follows from those of the corresponding $2D$ Galilean particles \eqref{2DGP}. We now consider enhanced Galilean and Carroll strings in $4D$. We will only consider those symmetries that are not realized manifestly at the level of the sigma model, see table 8.

 \begin{table}[!ht]
\renewcommand{\arraystretch}{2}
\begin{center}
\begin{tabular}{|c|c|c|}
\hline
{\bf Generator}&{\bf Field}&{\bf Parameter}\\
\hline
$J_{AB}, J_{ab}$&--&--\\
$G_{Aa}$&--&$\lambda^{Aa}$\\
$Z_a{}^b$ with $Z_a{}^a=0$&--&$\sigma^a{}_b$ with $\sigma^a{}_a=0$\\
$Z_A{}^B$ with $Z_A{}^A=0$&--&$\sigma^A{}_B$ with $\sigma^A{}_A=0$\\
$H_A, P_{a}$&$\tau^A, E^{a}$&g.c.t.\\
$M_{A}, M_a$&$m^{A}, n^a$&$\sigma^{A}, \sigma^a$\\
\hline
\end{tabular}
\end{center}
\caption{In this table we only give the parameters of the symmetries that are not realized manifestly in the sigma models given below.  We do not give the longitudinal and transverse rotations. Neither do we give  the general coordinate transformations (g.c.t.) which  are realized  in the sigma models as a so-called sigma model symmetry.}
\end{table}

\vskip .2truecm
 \noindent {\bf C. $\pmb{4D}$ Enhanced Galilei string}
 \vskip .2truecm
 The  Lagrangian of the $4D$ enhanced Galilei  string in a $4D$ string ($p=1$) enhanced Galilei  gravity background with field content $\{\tau_{\hat\mu}{}^{\hat A}, E_{\hat\mu}{}^{\hat a}, m_{\hat\mu}{}^{\hat A}\}$  is given by \cite{Bergshoeff:2018yvt}
 \begin{subequations}\label{4DGS}
\begin{align}
\mathcal{L}&=T\sqrt{h}h^{\alpha\beta}\partial_{\alpha}X^{\hat{\mu}}\partial_{\beta}X^{\hat{\nu}}
\Big[E_{\hat{\mu}}{}^{\hat{a}}E_{\hat{\nu}}{}^{\hat{b}}\delta_{\hat{a}\hat{b}}-2m_{(\hat{\mu}}{}^{\hat{A}}
\tau_{\hat{\nu})}{}^{\hat{B}}\eta_{\hat{A}\hat{B}}\Big]\label{4DGS1}\\[.1truecm]
&\textrm{with}\ \ h_{\alpha\beta}=\partial_{\alpha}X^{\hat{\mu}}\partial_{\beta}X^{\hat{\nu}}\tau_{\hat{\mu}}{}^{\hat{A}}
\tau_{\hat{\nu}}{}^{\hat{B}}\eta_{\hat{A}\hat{B}}\\[.1truecm]
&\textrm{and}\ \
D_{[\hat\mu}\tau_{\hat\nu]}{}^{\hat A} =0\ \textrm{(zero torsion)}\,,\label{4DGS3}
\end{align}
\end{subequations}
where we have integrated out the Lagrange multipliers. The curved hatted indices denote $4D$ indices: $\hat\mu = 0,1,2,3$ while the two flat indices $(\hat A,\hat a)$ refer to the two longitudinal and two transverse directions, respectively. The action  corresponding to the  $4D$ enhanced Galilei string \eqref{4DGS1} is invariant under the following Galilean boosts,  first and second non-central charge transformations  with parameters $\lambda^{\hat A\hat a}, \lambda^{\hat A}$ and  $\sigma^{\hat A}{}_{\hat B}\ (\textrm{with}\ \sigma^{\hat A}{}_{\hat A}=0)$ , respectively:\,\footnote{\label{footnote1} To show the invariance of the action corresponding to the Lagrangian \eqref{4DGS} under the first and second non-central charge transformation, it is useful to use the identities
\begin{equation}
h^{\alpha\beta} = \epsilon^{\alpha\gamma}\epsilon^{\beta\delta}h_{\gamma\delta}/h\,,\hskip 1.5truecm
\textrm{det}\, \tau_\alpha{}^A = \tfrac{1}{2}\epsilon^{\alpha\beta}\tau_\alpha{}^{\hat A}\tau_\beta{}^{\hat B}\epsilon_{\hat A\hat B}\,.
\end{equation}
Additionally, to prove the invariance under the first non-central charge transformation, one needs to use the zero torsion constraint \eqref{4DGS3}.
}
\begin{eqnarray}
\delta E_{\hat \mu}{}^{\hat a} &=& \lambda^{\hat a}{}_{\hat A}\, \tau_{\hat\mu}{}^{\hat A}\,,\\[.1truecm]
\delta m_{\hat \mu}{}^{\hat A} &=& D_{\hat \mu}\sigma{}^{\hat A} + \lambda^{\hat A}{}_{\hat a}\,E_{\hat\mu}{}^{\hat a} +
 \sigma^{\hat A}{}_{\hat B}\,\tau_{\hat \mu}{}^{\hat B}\,.
\end{eqnarray}
Note that the proof of invariance of the $4D$ enhanced Galilei string action under the above symmetries extends to any spacetime dimension. This is due to the fact that the range of the $\alpha$ and $\hat A$ indices do not change when going to higher dimensions and therefore the same identities as given in footnote \ref{footnote1}  can be used.

\vskip .2truecm
 \noindent {\bf D. $\pmb{4D}$ enhanced Carroll string}
 \vskip .2truecm

 Using the formal map between the $4D$ string ($p=1$) enhanced Galilei and enhanced Carroll algebras the Lagrangian \eqref{4DGS1} leads to the following corresponding Lagrangian for the $4D$ enhanced Carroll string in a $4D$ string ($p=1$) enhanced  Carroll gravity background with field content
$\{\tau_{\hat\mu}{}^{\hat A}, E_{\hat\mu}{}^{\hat a}, n_{\hat\mu}{}^{\hat a}\}$:

\begin{subequations}
\begin{align}
\mathcal{L}&=T\sqrt{h}h^{\alpha\beta}\partial_{\alpha}X^{\hat{\mu}}\partial_{\beta}X^{\hat{\nu}}
\Big[\tau_{\hat{\mu}}{}^{\hat{A}}\tau_{\hat{\nu}}{}^{\hat{B}}\eta_{\hat{A}\hat{B}}+2n_{(\hat{\mu}}{}^{\hat{a}}
E_{\hat{\nu})}{}^{\hat{b}}\delta_{\hat{a}\hat{b}}\Big]\\[.1truecm]
&\textrm{with}\ \ h_{\alpha\beta}=\partial_{\alpha}X^{\hat{\mu}}\partial_{\beta}X^{\hat{\nu}}E_{\hat{\mu}}{}^{\hat{a}}
E_{\hat{\nu}}{}^{\hat{b}}\delta_{\hat{a}\hat{b}}\\[.1truecm]
&\textrm{and}\ \
D_{[\hat\mu}E_{\hat\nu]}{}^{\hat a} =0\,.
\end{align}
\end{subequations}
The action  corresponding to this Lagrangian is invariant under the following Carroll boosts,  first and second non-central charge transformations  with parameters $\lambda^{\hat A\hat a}, \lambda^{\hat a}$ and  $\sigma^{\hat a}{}_{\hat b}\ (\textrm{with}\ \sigma^{\hat a}{}_{\hat a}=0)$ , respectively:
\begin{eqnarray}
\delta \tau_{\hat \mu}{}^{\hat A} &=& \lambda^{\hat A}{}_{\hat a}\, E_{\hat\mu}{}^{\hat a}\,,\\[.1truecm]
\delta n_{\hat \mu}{}^{\hat a} &=& D_{\hat \mu}\sigma{}^{\hat a} + \lambda^{\hat a}{}_{\hat A}\,\tau_{\hat\mu}{}^{\hat A} +
 \sigma^{\hat a}{}_{\hat b}\,E_{\hat \mu}{}^{\hat b}\,.
\end{eqnarray}
Unlike the enhanced Galilean string, the enhanced Carroll string is only invariant under the above symmetries in $4D$. This is related to the fact that the range of the index $\hat a$ is only 2 in  $4D$ and, according to footnote \ref{footnote1}, this is needed in the proof of invariance.

We now use the above string-string duality in $4D$ to see whether it leads to a corresponding   string-particle duality between sigma models in $3D$.  For this purpose, we will  perform below four different reductions leading to the following $3D$ particles and strings:

\begin{description}
\item {(1)} enhanced Galilei particle
\item{(2)} enhanced Galilei string
\item{(3)} enhanced Carroll string
\item{(4)} enhanced Carroll particle
\end{description}
Each of these four reductions requires a reduction of the corresponding background fields that are described, independent of their couplings to sigma models, in Appendix B.

\vskip .2truecm

\noindent {\bf (1) $\pmb{3D}$ enhanced Galilei particle}
\vskip .2truecm

We first reduce the $4D$ enhanced Galilean string over a spatial longitudinal direction. For the background fields we use the double dimensional reduction Ansatz \eqref{doubleGalilean}-\eqref{doubleGalilean2}  to go from $4D$ enhanced Galilei gravity  to $3D$ enhanced Galilei gravity with field content $\{\tau_\mu, E_\mu{}^a,m_\mu\}$ as given in Appendix A.

In this so-called double dimensional reduction of the string we have imposed the gauge
 $X^{y}=\sigma$ where $(t,\sigma)$ are the worldsheet coordinates. We have furthermore assumed that all background fields are independent of $X^y=\sigma$.
After reduction we obtain the Lagrangian for a $3D$ enhanced Galilei particle:
\begin{subequations}\label{3DBGparticle}
\begin{align}
\mathcal{L}&=m e^{-1}\dot{X}^{\mu}\dot{X}^{\nu}\Big[E_{\mu}{}^{a}E_{\nu}{}^{b}\delta_{ab}-2m_{(\mu}\tau_{\nu)}\Big]\\
e&=\dot{X}^{\mu}\tau_{\mu}\ \ \ \textrm{and}\ \ \ \partial_{[\mu}\tau_{\nu]}=0\, ,
\end{align}
\end{subequations}
where the string tension $T$ becomes $m^2$ under the reduction.
The action corresponding to this Lagrangian is invariant under the following Galilean boosts and central charge transformations with parameters $\lambda^a$ and $\lambda$, respectively:
\begin{eqnarray}
\delta E_\mu{}^a &=& \lambda^a\tau_\mu\,,\\[.1truecm]
\delta m_\mu &=& \partial_\mu\sigma + \lambda^a E_\mu{}^b\,\delta_{ab}\,.
\end{eqnarray}
\vskip .2truecm

\noindent {\bf (2) $\pmb{3D}$ enhanced Galilei string}
\vskip .2truecm
 Performing   a direct dimensional  reduction of  4D string ($p=1$) enhanced Galilei  gravity  to 3D string ($p=1$) enhanced Galilei gravity with field content $\{\tau_\mu{}^A, E_\mu, m_\mu{}^A\}$ as given in
 eqs.~\eqref{directGalilean} - \eqref{directGalilean2} of Appendix A,
we obtain the following  Lagrangian for a $3D$ enhanced Galilei string in a $3D$ string ($p=1$) enhanced Galilei  background:
\begin{subequations}\label{3DBGstring}
\begin{align}
\mathcal{L}&=T\sqrt{h}h^{\alpha\beta}\partial_{\alpha}X^{\mu}\partial_{\beta}X^{\nu}\Big[E_{\mu}E_{\nu}+2m_{(\mu}{}^{A}\tau_{\nu)}{}^{B}
\eta_{AB}\Big]\\
h_{\alpha\beta}&=\partial_{\alpha}X^{\mu}\partial_{\beta}X^{\nu}\tau_{\mu}{}^A\tau_{\nu}{}^B\eta_{AB}\ \
\textrm{and}\ \ D_{[\mu}\tau_{\nu]}{}^A=0\,.
\end{align}
\end{subequations}
Note that we have also truncated away in a consistent way the transverse scalar $X^y$.
The action  corresponding to this Lagrangian is invariant under the following Galilean boosts, first and second central charge transformations with parameters $\lambda^A, \sigma^A$ and $\sigma^A{}_B$ with $\sigma^A{}_A=0$, respectively:
\begin{eqnarray}
\delta E_\mu &=& \lambda^A\, \tau_\mu{}^B\, \eta_{AB}\,,\\[.1truecm]
\delta m_\mu{}^A &=& D_\mu\sigma^A + \lambda^A\,E_\mu + \sigma^A{}_B \tau_\mu{}^B\,.
\end{eqnarray}
\vskip .2truecm

We now consider the two different reductions of the $4D$ Carroll string. Here  things are more complicated due to the fact that a naive Ansatz leads to a degenerate world-sheet metric. As we will see below, to avoid such a singular metric,  in the case of the Carroll string the number of transverse directions has to be larger or equal to two whereas
in the case of the Carroll particle   we should keep several matter fields when reducing the background Carroll fields.
\vskip .2truecm

\noindent {\bf (3)  $\pmb{3D}$ enhanced Carroll string}
\vskip .2truecm

At first sight, one would expect that  the $3D$ enhanced Galilei particle discussed in (1) is formally dual to a $3D$ enhanced Carroll string. From the $4D$ point of view this means that, when we reduce the enhanced Galilei string over a longitudinal direction, we should instead reduce the corresponding $4D$ enhanced Carroll string over a transverse direction.

To  obtain the enhanced Carroll string, we perform a direct dimensional reduction of $4D$ string ($p=1$) enhanced Carroll gravity  to $3D$ string ($p=1$) enhanced Carroll  gravity with field content $\{\tau_\mu{}^A, E_\mu\,, n_\mu\}$ as given in \eqref{directCarroll} - \eqref{directCarroll2} of Appendix A.
After reduction we obtain the following Lagrangian for the  $3D$ enhanced Carroll string:
\begin{subequations}
\begin{align}
\mathcal{L}&=T\sqrt{h}h^{\alpha\beta}\partial_{\alpha}X^{\mu}\partial_{\beta}X^{\nu}\Big[
\tau_{\mu}{}^{A}\tau_{\nu}{}^{B}\eta_{AB}+2n_{(\mu}E_{\nu)}\Big]\label{3DBCstring}\\
h_{\alpha\beta}&=\partial_{\alpha}X^{\mu}\partial_{\beta}X^{\nu}E_{\mu}E_{\nu} + \partial_\alpha X^y \partial_\beta X^y\ \ \ \textrm{and}\ \ \ \partial_{[\mu}E_{\nu]}=0\,.
\end{align}
\end{subequations}
\noindent Note that, unlike the Galilei string, we cannot truncate away in a consistent way the transverse scalar $X^y$ since this would lead to a degenerate world-sheet metric. This has the effect that our answer for the $3D$  enhanced Carroll string   is not formally dual to that of the Galilei particle.
The action corresponding to the Lagrangian \eqref{3DBCstring} is invariant under the following Carroll boosts and central charge transformations with parameters $\lambda^A$ and $\lambda$, respectively:
\begin{eqnarray}
\delta \tau_\mu{}^A &=& \lambda^A\,E_\mu\,,\\[.1truecm]
\delta n_\mu &=& \partial_\mu\sigma - \lambda^A \tau_\mu{}^B\,\eta_{AB}\,.
\end{eqnarray}

\vskip .2truecm

\noindent {\bf (4) $\pmb{3D}$ enhanced Carroll particle}
\vskip .2truecm
 We first impose the gauge-fixing condition
 \begin{align}
X^{y}=\sigma
\end{align}
and assume that nothing depends on $X^{y}$. Next, we observe that the reduction of the worldsheet metric gives
\begin{align}
h_{\alpha\beta}&=\left(\begin{array}{cc}
E^{a}E_{a}&E^{a}E_{y a}\\
E^{a}E_{y a}&E^{a}_{y}E_{y a}
\end{array}\right)\,,
\end{align}
where we have defined
\begin{align}
E^{a}&=\partial_{\tau}X^{\mu}E_{\mu}^{a}\,.
\end{align}
This implies that  $h_{\alpha\beta}$ is degenerate if we set the scalar fields $E_{y}^{a}$ to zero. Therefore, we need to work with matter-coupled Carroll gravity. Keeping these scalars we write the Carroll string action as
\begin{subequations}
\begin{align}
\mathcal{L}&=T\sqrt{h}h^{\alpha\beta}\partial_{\alpha}X^{\hat{\mu}}\partial_{\beta}X^{\hat{\nu}}
\bigg[\tau_{\hat{\mu}}^{\hat{A}}\tau_{\hat{\nu}}^{\hat{B}}\eta_{\hat{A}\hat{B}}+2n_{(\hat{\mu}}^{\hat{a}}E_{\hat{\nu})}^{\hat{b}}
\delta_{\hat{a}\hat{b}}\bigg]\\
&=T\sqrt{h}h^{\alpha\beta}\tau_{\alpha}^{\hat{A}}\tau_{\beta}^{\hat{B}}\eta_{\hat{A}\hat{B}}+2\epsilon_{\hat{a}\hat{b}}
\epsilon^{\alpha\beta}E_{\alpha}^{\hat{a}}n_{\beta}^{\hat{b}}\,,
\end{align}
\end{subequations}
 where
  \begin{align}
h=E^{2}E_{y}^{2}-(EE_{y})^2=(\epsilon_{ab}E^{a}E_{y}^{b})^2\,.
\end{align}

 Performing a double dimensional reduction of $4D$ string enhanced Carroll gravity to $3D$ particle enhanced Carroll gravity with field content $\{\tau_\mu\,, E_\mu{}^a\,, n_\mu{}^a\}$ coupled to matter with field content $\{\tau_y\,, E_y{}^a\,, n_y{}^a\}$
 as given in eqs.~\eqref{doubleCarroll} - \eqref{doubleCarroll2} of Appendix A,
 we obtain the  following Lagrangian for a $3D$ enhanced Carroll particle in a $3D$ enhanced matter-coupled Carroll gravity background:
\begin{align}
\mathcal{L}&=-Te^{-1}(E_{y}^{a}\tau-E^{a}\tau_{y})(E_{y a}\tau-E_{a}\tau_{y})+2T\epsilon_{ab}(E^{a}n_{y}^{b}-E^{a}_{y}n^{b})\\[.1truecm]
&\textrm{with}\ \  e^{-1}=\epsilon_{ab}E^{a}E_{y}^{b}\hskip .5truecm \textrm{and}\hskip .5truecm
D_{[\mu}E_{\nu]}^{a}=
D_{\mu}E_{y}^{a}=0\,.
\end{align}
 The action corresponding to this Lagrangian is invariant under the following transformation rules:
\begin{subequations}
\begin{align}
\delta \tau_{\mu}&=\lambda{}_{a}E_{\mu}{}^{a}\,,\\[.1truecm]
\delta \tau_{y}&=\lambda{}_{a}E_{y}{}^{a}\,,\\[.1truecm]
\delta n_{\mu}^{a}&=D_{\mu}\sigma^{a}+\lambda^{a}\tau_{\mu}+\sigma^{a}{}_{b}E_{\mu}{}^{b}\,,\\[.1truecm]
\delta n_{y}^{a}&=\lambda^{a}\tau_{y}+\sigma^{a}{}_{b}E_{y}{}^{b}\,.
\end{align}
\end{subequations}
All background fields are needed to avoid a singular Carroll string in $4D$ and to keep the boost invariance.

This finishes our discussion of enhanced Carroll particles and strings in three dimensions.

\section{Conclusions}

In this work we exploited a formal relation between Galilei and Carroll symmetries, first observed in \cite{Barducci:2018wuj}, to obtain new results on Carroll gravity theories and the couplings of particles and strings to a curved Carroll background by making use of known results in the Galilei case. A crucial ingredient was to consider these symmetries from a brane perspective instead of particles only. We pointed out that the formal relationship between Galilei and Carroll remained valid in the Lie algebra expansion of the Poincar\'e algebra. This enabled us to construct several new examples of Carroll gravity actions. Concerning the sigma model couplings, we found both example without (non-) central charge symmetries and including (non-)central charge symmetries. The first category corresponds to massless representations of the Galilei and Carroll algebras, which have been discussed before in the literature. The second category leads to several examples of massive particle and string actions in two, three and four dimensions. We made use here of a particle-particle duality in two dimensions and a string-string duality in four dimensions. Other examples of sigma models were derived from these cases by dimensional reduction but the results are not so symmetric anymore under the Galilei-Carroll symmetry discussed in this paper. We related this to the fact that the naive dimensional reduction leads to a degenerate world-sheet metric that could be avoided by either keeping the extra higher-dimensional embedding scalar ($3D$ Carroll strings) or by working with matter coupled Carroll gravity ($3D$ Carroll particle).

Extending this method to higher dimensions one could consider  membrane-membrane dualities in six spacetime dimensions. The latter should lead, after dimensional reduction, not only to string-membrane  dualities in $5D$ but also to string-string and particle-membrane dualities in $4D$. This should lead to many more examples of branes moving in a curved Galilei or Carroll background.

It would be interesting to extend the methods exploited in this paper to other algebras such as the AdS algebra including a cosmological constant or supersymmetric algebras involving fermionic fields. We hope to come back to these algebras in the nearby future.

\FloatBarrier
\section*{Acknowledgements}

We thank Johannes Lahnsteiner, Jan Rosseel and Ceyda \c Sim\c sek  for discussions and Quim Gomis for useful comments on the manuscript. In particular, we wish to thank Tomas Ort\'{i}n who was involved in the initial stage of this project. Part of this work was done while E.A.B. was visiting the Instituto de F\'{\i}sica Te\'orica (IFT) of the UAM/CSIC in Madrid as a  IFT Severo Ochoa Associated Researcher. E.A.B. wishes to thank the IFT for its hospitality.

\appendix
\section{Nomenclature}

In order to avoid confusion with the many different gravity theories that circulate in this paper, we have introduced the following nomenclature thereby avoiding using the word Bargmann since that name is too specific:

\begin{description}

\item {(1)} The gravity theories corresponding to the smallest algebras  that occur at lowest order in the Lie algebra expansion of the Poincar\'e algebra are called Galilei and Carroll gravity. The algebras can be obtained by a In\"on\"u-Wigner  contraction of the Poincaré algebra and the dynamics of the gravity theories can be described by an action.

\item{(2)} The smallest extensions of the Galilei and Carroll gravity theories corresponding to an algebra  that does ${\underline{\rm not}}$  occur in the Lie algebra expansion of the Poincar\'e algebra are called the {\sl enhanced} Galilei and the {\sl enhanced} Carroll algebra. Using this nomenclature, we call the Bargmann algebra the enhanced Galilei algebra. These are the gravity theories that can be coupled to particles and/or strings but we cannot write down an action for them.

\item{(3)} The smallest extensions  of  the Galilei and Carroll gravity theories corresponding to an algebra that occurs  in the Lie algebra expansion of the Poincar\'e algebra and  for which we can write down an action are called the {\sl extended} Galilei algebra and the {\sl extended} Carroll algebra. Using this nomenclature we call the $3D$ extended Bargmann algebra the $3D$ extended Galilei algebra.
The dynamics of these gravity theories is described by an action but in general we do not know how to couple these gravity theories  to particles and/or strings.\footnote{In low dimensions there could be exceptions, due to the fact that some generators simply do not appear; an example of this is represented by the 3D enhanced string Galilei algebra of Table 7 when $Z_{AB}$ is restricted to its antisymmetric part. This algebra could both occur in sigma models and in the Lie algebra expansion.} A prime example is the extended Bargmann algebra leading to extended Bargmann gravity. An action for extended Bargmann gravity has been constructed \cite{Papageorgiou:2009zc,Bergshoeff:2016lwr,Hartong:2016yrf} but a coupling to a particle including the second central charge is not known.

\end{description}

For each of the above algebras we can distinguish between the particle ($p=0$) version, the string ($p=1$) version, the membrane ($p=2$) version etc., where the number $p$ refers to the number of longitudinal directions.  For the convenience of the reader we have summarized the main properties of the above gravity theories in Table 9.

\begin{table}[!ht]
\renewcommand{\arraystretch}{2}
\begin{center}
\begin{tabular}{|c|c|c|}
\hline
{\bf Name}&{\bf Coupling}&{\bf Action}\\
\hline
Galilei/Carroll&Yes&Yes\\
enhanced Galilei/Carroll&Yes&No\\
extended Galilei/Carroll&No&Yes\\
\hline
\end{tabular}
\end{center}
\caption{This Table summarizes the main properties of the enhanced/extended Galilei/Carroll gravity theories discussed in the work. By coupling in the second column we mean the coupling to particles and/or strings.}
\end{table}

\section{Dimensional Reduction}

In this appendix we discuss the dimensional reduction of the $4D$  enhanced string Galilei and Carroll  gravity theories  to three dimensions. Due to the foliated geometry one can either perform a so-called {\it double dimensional reduction} where the compactification direction coincides with one of the foliation directions, or one may perform a {\it direct dimensional reduction} where the compactification direction differs from the foliation directions. This leads to four possibilities for the dimensional reduction of the string ($p=1$) enhanced Galilei and Carroll gravity fields  which we shortly discuss one by one below.

To describe the different reductions, we will for simplicity omit the $(0)$ upper index for the lowest-order fields in the Lie algebra expansion used in sections 2 and 3. Furthermore, we will always indicate the second-order vector field, with upper index $(2)$, that occurs in the Galilei case with a $m_\mu$ and the one that occurs in the Carroll case with a $n_\mu$. Both 1-forms could have an additional other index in some cases. Finally, we will indicate all $4D$ indices with a hat and decompose them as $\hat\mu = (\mu,y)$ and $(\hat A, \hat a)  = (A,a)
= (0,1,2,3)$.

\vskip .5truecm

\noindent {\bf (1)\ \ Double dimensional reduction of $4D$ enhanced string Galilei gravity}
\vskip .2truecm

The double dimensional reduction of $4D$ string ($p=1$) enhanced Galilei  gravity with field content $\{\tau_{\hat\mu}{}^{\hat A}\,,  E_{\hat\mu}{}^{\hat a}\,, m_{\hat \mu}{}^{\hat A}\}$ to $3D$ particle ($p=0$) enhanced Galilei  gravity with field content $\{\tau_\mu, E_\mu{}^a\,, m_\mu\}$ is given by
\begin{multieq}[3]\label{doubleGalilean}
E_{\mu}^{\hat{a}}&=E_{\mu}^{a}\\
E_{y}^{\hat a}&=0\\
m_{\mu}^{1}&=0\\
\tau_{y}^{0}&=0\\
m_{y}^{0}&=0\\
m_{y}^{1}&=0\\
\tau_{\mu}^{1}&=0\\
\tau_{y}^{1}&=1\\
m_{\mu}^{0}&=m_{\mu}\\
\tau_{\mu}^{0}&=\tau_{\mu}\label{doubleGalilean2}
\end{multieq}
\noindent

\vskip .5truecm

\noindent {\bf (2)\ \ Direct dimensional reduction of $4D$ enhanced string Galilei gravity}
\vskip .2truecm

The direct  dimensional reduction of $4D$  string ($p=1$) enhanced Galilei  gravity with field content $\{\tau_{\hat\mu}{}^{\hat A}\,,  E_{\hat\mu}{}^{\hat a}\,, m_{\hat \mu}{}^{\hat A}\}$ to $3D$ string ($p=1$) enhanced  Galilei  gravity with field content $\{\tau_\mu{}^A, E_\mu\,, m_\mu{}^A\}$ is given by
\begin{multieq}[3]\label{directGalilean}
m_{\mu}^{\hat{A}}&=m_{\mu}^{A}\\
\tau_{\mu}^{\hat{A}}&=\tau_{\mu}^{A}\\
E_{\mu}^{2}&=E_{\mu}\\
E_{\mu}^{3}&=0\\
E_{y}^{3}&=1\\
E_{y}^{2}&=0\\
\tau_{y}^{\hat{A}}&=0\\
m_{y}^{\hat{A}}&=0\label{directGalilean2}
\end{multieq}

\vskip .5truecm

\noindent {\bf (3)\ \ Direct dimensional reduction of $4D$ enhanced string  Carroll gravity}
\vskip .2truecm

 The  direct dimensional reduction of  $4D$ enhanced string Carroll gravity with field content $\{\tau_{\hat\mu}{}^{\hat A}, E_{\hat\mu}{}^{\hat a}, n_{\hat\mu}{}^{\hat a}\}$ to $3D$ enhanced string Carroll gravity with field content $\{\tau_\mu{}^A, E_\mu, n_\mu\}$ is given by:
\begin{multieq}[3]\label{directCarroll}
\tau_{\mu}^{\hat{A}}&=\tau_{\mu}^{A}\\
\tau_{y}^{\hat{A}}&=0\\
n_{\mu}^{2}&=n_{\mu}\\
E_{\mu}^{3}&=0\\
E_{\mu}^{2}&=E_{\mu}\\
n_{y}^{2}&=0\\
E_{y}^{2}&=0\\
n_{\mu}^{3}&=0\\
E_{y}^{3}&=1\\
n_{y}^{3}&=0\label{directCarroll2}
\end{multieq}

\noindent {\bf (4)\ \ Double dimensional reduction of $4D$ enhanced string  Carroll gravity}
\vskip .2truecm

The double dimensional  reduction of $4D$ enhanced  string Carroll gravity with field content $\{\tau_{\hat\mu}{}^{\hat A}, E_{\hat\mu}{}^{\hat a}, n_{\hat\mu}{}^{\hat a}\}$ to $3D$ enhanced Carroll gravity with field content
$\{\tau_\mu\,, E_\mu{}^a\,, n_\mu{}^a\}$ coupled to matter with field content $\{\tau_y\,, E_y{}^a\,, n_y{}^a\}$ is given by
\begin{multieq}[3]\label{doubleCarroll}
n_{\mu}{}^{\hat{a}}&=n_{\mu}{}^{a}\\
E_{\mu}{}^{\hat{a}}&=E_{\mu}{}^{a}\\
\tau_{\mu}^{0}&=\tau_{\mu}\\
\tau_{\mu}^{1}&=0\\
\tau_{y}^{0}&=\tau_{y}\\
\tau_{y}^{1}&=0\\
E_{y}{}^{\hat{a}}&=E_{y}{}^{a}\\
n_{y}{}^{\hat{a}}&=n_{y}{}^{a}\label{doubleCarroll2}
\end{multieq}
When coupling to a Carroll particle the matter cannot be truncated away in a consistent way and/or  without giving up boost invariance, see the main text.

\FloatBarrier

\end{document}